\documentclass[twocolumn,titlepage,preprintnumbers,amsmath,amssymb,article.cls,superscriptaddress]{revtex4-1}
\usepackage{textcomp} 
\usepackage[caption=false]{subfig} 
\usepackage{enumerate}
\usepackage{paralist}
\usepackage{ulem}
\usepackage{eqnarray}
\usepackage{subeqnarray}
\usepackage{amsmath}
\usepackage{amssymb,amsfonts,latexsym}
\usepackage{bm}
\usepackage[mathcal]{euscript}
\usepackage{graphicx}
\usepackage{epsfig}
\usepackage{color}
\usepackage{pifont}
\definecolor{blue}{rgb}{0,0,1}
\definecolor{bleuf}{rgb}{0,0,0.9}
\definecolor{rougef}{rgb}{0.9,0,0}
\definecolor{green}{rgb}{0,0.5,0}
\definecolor{red}{rgb}{1,0,0}
\definecolor{pink}{rgb}{0.9,0.3,0.7}
\definecolor{azur}{rgb}{0,0.5,0.5}
\definecolor{orange}{rgb}{1,0.5,0.2}
\definecolor{brown}{rgb}{0.5,0,0}

\newcommand{\be}{\begin{equation}}
\newcommand{\ee}{\end{equation}}
\newcommand{\ben}{\begin{equation*}}
\newcommand{\een}{\end{equation*}}
\newcommand{\ba}{\begin{eqnarray}}
\newcommand{\ea}{\end{eqnarray}}

\begin{document}
\title{\textbf{Self-sustained lift and low friction via soft lubrication}}
\author{Baudouin Saintyves}
\affiliation{Paulson School of Engineering and Applied Sciences, Harvard University, Cambridge, MA 02138, USA}
\author{Th\'eo Jules \footnote{The two first authors contributed equally.}}
\affiliation{Paulson School of Engineering and Applied Sciences, Harvard University, Cambridge, MA 02138, USA}
\affiliation{D\'epartement de Physique, ENS, PSL Research University, 75005 Paris, France}
\author{Thomas Salez}
\affiliation{Paulson School of Engineering and Applied Sciences, Harvard University, Cambridge, MA 02138, USA}
\affiliation{PCT Lab, UMR CNRS 7083 Gulliver, ESPCI Paris, PSL Research University, 75005 Paris, France}
\author{L. Mahadevan}
\email{ lm@seas.harvard.edu}
\affiliation{Paulson School of Engineering and Applied Sciences, Department of Physics, Wyss Institute for Bioinspired Engineering and the Kavli Institute for Nanobio Science and Technology, Harvard University, Cambridge, MA 02138, USA}
\date{\today}

\begin{abstract}
Relative motion between soft wet solids arises in a number of applications in natural and artificial settings, and invariably couples elastic deformation and fluid flow. We explore this in a minimal setting by considering a fluid-immersed negatively-buoyant cylinder moving along a soft inclined wall. Our experiments show that there is an emergent robust steady-state sliding regime of the cylinder with an effective friction that is significantly reduced relative to that of rigid fluid-lubricated contacts. A simple scaling approach that couples the cylinder-induced flow to substrate deformation allows us to explain the emergence of an elastohydrodynamic lift that underlies the self-sustained lubricated motion of the cylinder, consistent with recent theoretical predictions. Our results suggest an explanation for a range of effects such as reduced wear in animal joints and long-runout landslides, and can be couched as a design principle for low-friction interfaces.
\end{abstract}

\maketitle

Sliding motion between contacting solids arises in a range of phenomena that spans many length and time scales and includes landslides~\cite{Campbell1989}, aquaplaning of tires~\cite{Brochard2003}, industrial bearings~\cite{Hamrock1994}, synovial and cartilaginous joints~\cite{Maroudas76,Greene2011,Grodzinsky1978,Mow2002,Mow1984}, cell motion in blood vessels and microfluidic devices~\cite{Goldsmith1971,Byun2013} and atomic-force and surface-force rheological apparati~\cite{Villey2013}. Interfacial sliding invariably involves friction and adhesion, as well as fluid lubrication and elastic deformation \cite{Persson, Persson2009}.  

Since the pioneering work of Reynolds~\cite{Reynolds1886}, fluid lubrication has been extensively studied, initially in the context of heavy industry~\cite{Hamrock1994}, and more recently in the context of motion at soft material and biological interfaces~\cite{Zeng2013}. In heavy load, high velocity settings, strongly confined induced viscous flows can generate high pressure and heat, with associated rheological piezoviscous, thermoviscous effects and nanometric substrate deformations~\cite{Hamrock1994}. In contrast, in light loading conditions associated with steady sliding motions at soft wet interfaces, the coupling between elasticity and flow leads to long wavelength deformations that predict the emergence of lift and reduced friction~\cite{Sekimoto1993,Skotheim2004,Skotheim2005,Snoeijer2013}. Perhaps surprisingly, it is only recently that the problem of a free particle that can simultaneously sediment, slide or roll has been treated~\cite{Salez2015}, with further predictions of a range of counterintuitive solutions such as enhanced sedimentation, bouncing, roll reversal and self-sustained long-runout sliding. Here we consider a minimal experimental setting to test these predictions, and focus in particular on the self-sustained elastohydrodynamic lift and the accompanying low effective friction associated with the motion of a heavy fluid-immersed cylinder sliding along a soft inclined wall.   

\begin{figure}[h]
\begin{center} 
\includegraphics[width=0.45\textwidth]{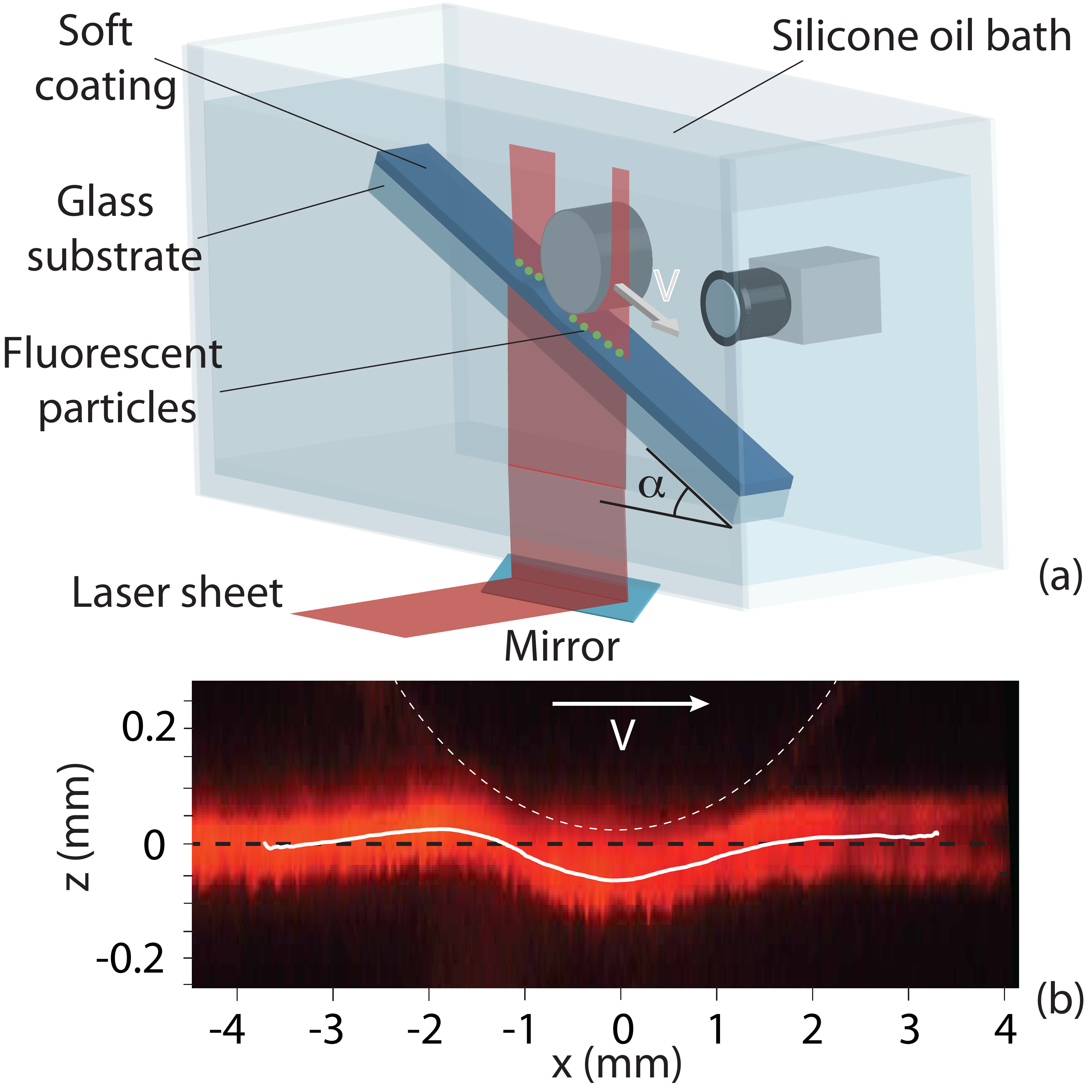}
\end{center}
\caption{\textbf{Experimental setup and laser profilometry.} (a) A negatively-buoyant rigid cylinder immersed in a viscous bath slides down a tilted wall that is coated with a thin elastic layer. Fluorescent particles embedded in the latter allow for laser profilometry {of the elastic layer-oil interface}. (b) Side view of the substrate deformation (red) obtained with the laser sheet. The centre of the cylinder is located at $x=0$. The white dashed line is a guide to the eye indicating the cylinder profile, the black dashed line corresponds to the unperturbed state of the substrate. The white solid line corresponds to the center of the fluorescent signal from the particles, obtained by using a transverse Gaussian fit, showing the asymmetric profile of the elastic layer-oil interface. The experimental parameters are $G=65$~kPa, $h=1.5$~mm, $\eta=1$~Pa.s, $R=12.7$~mm, {$\rho=8510$~kg/m$^3$}, $\alpha=11^{\circ}$.\label{fig1}} 

\end{figure}

Our setup, sketched in Fig.~\ref{fig1}(a), consists of a glass plate ($0.5~\textrm{cm}\times5~\textrm{cm}\times30~\textrm{cm}$) mounted on a variable incline (angle $\alpha$), coated with a thin layer of soft material (polyacrylamide-PAA or polydimethylsiloxane-PDMS with varying crosslinker and monomer concentrations, see SI) of thickness $h$ and shear modulus $G$, immersed in a transparent aquarium ($8~\textrm{cm}\times15~\textrm{cm}\times33~\textrm{cm}$) filled with Rhodorsil silicone oil of density $\rho_{\textrm{oil}}=970~\textrm{kg}/\textrm{m}^3$ and with varying viscosities $\eta$. Cylinders of radii $R$, made of different materials of density $\rho$ (relative density $\rho^*=\rho-\rho_{\textrm{oil}}>0$), are allowed to slide down the incline after being launched manually, and their trajectory is followed using a simple wide-angle camera. To visualize the profile of the deformed coating, we embed fluorescent polyethylene particles at the elastic-oil interface, use a blue laser sheet to illuminate the interface from below, and focus a near-field macro-lens on the interface, as shown in Fig.~\ref{fig1}(a); this allows us to see only the interface (Fig.~\ref{fig1}(b), Movie M1). In a given experiment, we observe both spinning and sliding motions of the cylinder (Movie M2), so that the translation speed is $\dot{\theta} R + V$, where $\dot{\theta}$ is the angular velocity and $V$ the sliding speed. For most experiments, $V \gg R\dot{\theta}$ (except for motion near very soft substrates, with $G<1000$~Pa) and thus we focus on the sliding motion that is extracted via an image-processing algorithm.

\begin{figure}[b]
\begin{center} 
\includegraphics[width=0.45\textwidth]{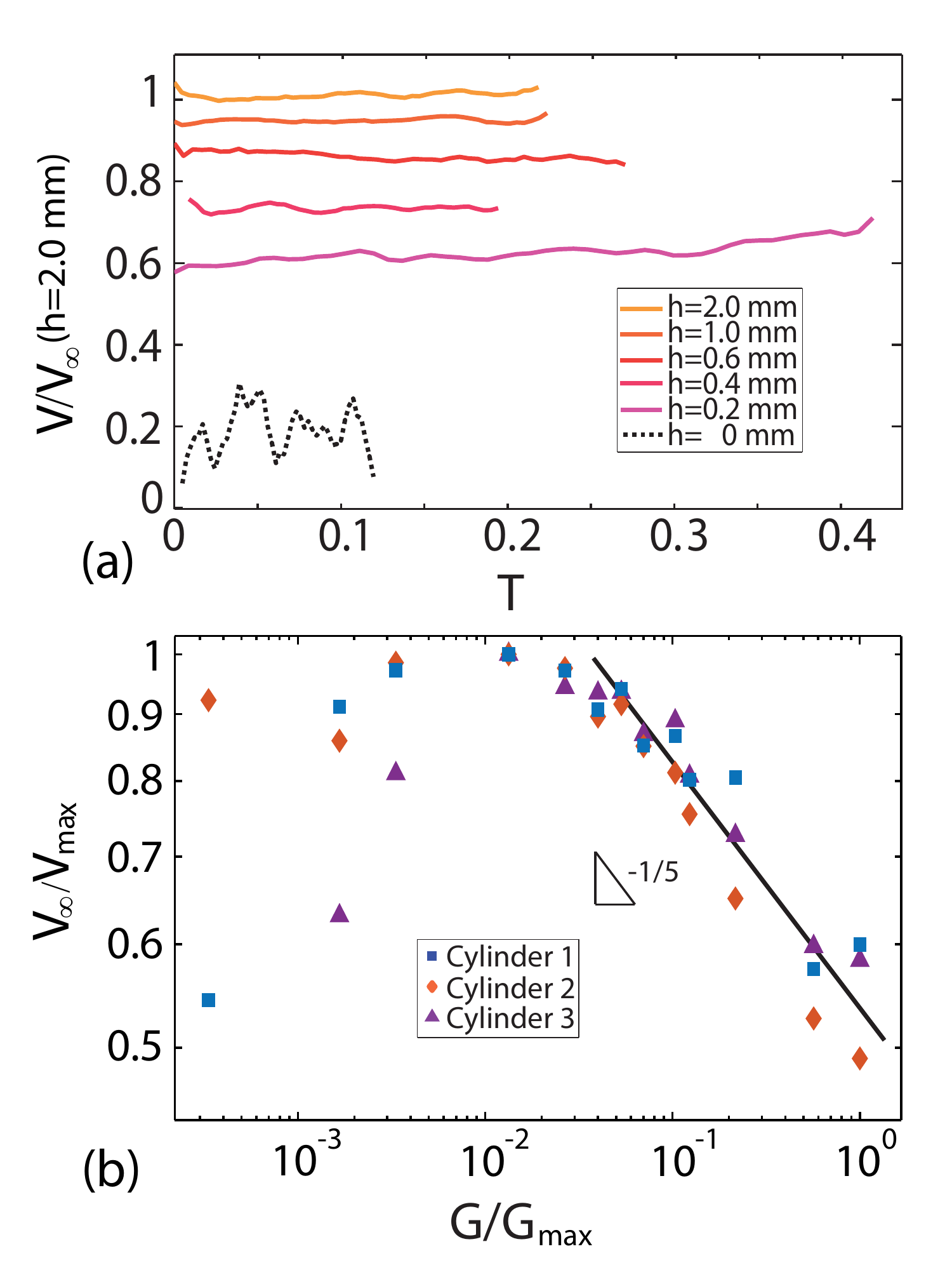}
\end{center}
\caption{\textbf{Sliding speed as a function of time and shear modulus of the substrate.} (a) Sliding speed $V$ of an aluminium cylinder as a function of normalized time $T=tV_{\infty}/d$, for different thicknesses $h$ of the soft coating. Here $V_{\infty}$ is the time-averaged steady-state sliding speed, $d$ the total length of the substrate, and $t$ is the time. The vertical axis is normalized by $V_\infty$ (for $h=0.6$~mm). The dotted line corresponds to the case of a bare glass substrate, while the solid lines are for coatings with an elastic layer of shear modulus $G=31$~kPa. The other experimental parameters are $\rho=2720$~kg/m$^3$, $R=12.7$~mm, $\eta=1$~Pa.s, and $\alpha=11^{\circ}$. (b) Time-averaged steady-state sliding speed for 3 different cylinders as a function of the shear modulus of the coating, with thickness $h=600\,\mu$m, oil bath viscosity $\eta=1$~Pa.s, and angle $\alpha=11^{\circ}$. The vertical axis is normalized by the maximum steady-state sliding speed $V_{\textrm{max}}$ for the corresponding cylinder. The horizontal axis is normalized by the maximum of the tested moduli $G_{\textrm{max}} = 300$~kPa. The solid line has a slope of $-1/5$ (Eq.~(\ref{master})). The other parameters are as follows. Cylinder 1 (aluminium): $\rho=2720$~kg/m$^3$, $R=12.7$~mm; cylinder 2 (glass): $\rho=2240$~kg/m$^3$, $R=9.5$~mm; cylinder 3 (brass): $\rho=8510$~kg/m$^3$, $R=6.35$~mm.\label{fig2}}
\end{figure}

For a base line, we first perform experiments with the cylinder moving down a rigid glass incline, and see stick-slip motion dominated by surface roughness~\cite{Bowden73}, rather than fluid-lubricated smooth motion of the cylinder considered theoretically~\cite{Brenner1962,Jeffrey1981}; the dotted line in Fig.~\ref{fig2}(a) highlights this intermittent low-speed motion that is very sensitive to initial conditions. When the cylinder moves on a soft coating, its motion transitions to that dominated by fluid-lubricated contact and there are a number of new effects. First, we observe that for certain combinations of cylinder size and density, the cylinder reverses its rolling direction (Movie M3), consistent with recent prediction~\cite{Salez2015}. Second, we also observe damped oscillations (Movie M4) when the cylinder is launched from a large height and it settles into its steady-state sliding motion, as predicted in~\cite{Salez2015}. Most strikingly, over a robust range of parameters, we observe a steady sliding of the cylinder accompanied by an asymmetrically deformed elastic-oil interface (Movie M1), as shown in Fig.~\ref{fig1}(b). This counter-intuitive observation suggests a mechanism for self-sustained lift at soft wet interfaces with implications for a range of phenomena that go beyond the simple instantiation here. 

It can be qualitatively understood by a simple modification of classical lubrication theory~\cite{Reynolds1886,Batchelor1967}. For the case of rigid interfaces, the fore-aft symmetry of the gap implies that the pressure field is antisymmetric with a null normal resultant force. However, when the wall or cylinder is soft, this pressure asymmetry leads to an asymmetric elastic deformation~\cite{Sekimoto1993,Skotheim2004,Skotheim2005,Salez2015} that generates lift dynamically which is capable of sustaining the cylinder's weight. Consistent with this, independent of initial conditions, over the entire range of tested substrate thicknesses and moduli, the cylinder achieves a robust sliding steady state (Fig.~\ref{fig2}(a)) with a speed that is substantially higher than its speed when sliding along a rigid glass wall. The same behaviour is also qualitatively observed for elliptical cylinders (SI, Movie M5) -- since these do not spin, we can disregard rotation as the primary origin of self-sustained lift and reduced friction. Moving to quantify the role of the soft substrate, we note that the thicker (Fig.~\ref{fig2}(a)) and the softer (Fig.~\ref{fig2}(b)) the elastic layer, the higher the sliding speed, up to a certain point. In Fig.~\ref{fig2}(b) we see that the sliding speed is a non-monotonic function of the modulus; it falls off for large moduli as expected, but also for small moduli owing to effects such as large substrate deformations (see Movie M6 for an example of an experiment with a very soft coating).

To minimally quantify the emergence of self-sustained lift and sliding motion of the cylinder at speed $V$, we limit ourselves to considerations of small strains in the elastic layer; as we will see, this suffices to explain most of our experiments. Assuming that the cylinder of radius $R$ is separated from the undeformed substrate by a minimum gap $\delta \ll R$, the tangential size of the contact zone scales as $l\sim\sqrt{R\delta}$, so that $\delta\ll l\ll R$. This separation of scales allows us to invoke lubrication theory~\cite{Batchelor1967}. Then, the flow-induced pressure $p$ scales as $p\sim\eta Vl/(\delta+\Delta h)^2$, where $\Delta h\sim h\,p/G\ll h$ is the normal elastic deformation of the soft layer~\cite{Johnson1985,Skotheim2004}. Together, this implies that the elastohydrodynamic pressure scales as $\eta^2 V^2Rh/(G \delta^4)$ at leading order. Integrating this pressure over the contact length $l$ leads to a positive elastohydrodynamic lift force per unit length that scales as $\sim \eta^2V^2R^{3/2}h/(G\delta^{7/2})$~\cite{Skotheim2004,Skotheim2005}.

\begin{figure}[b]
\begin{center} 
\includegraphics[width=0.45\textwidth]{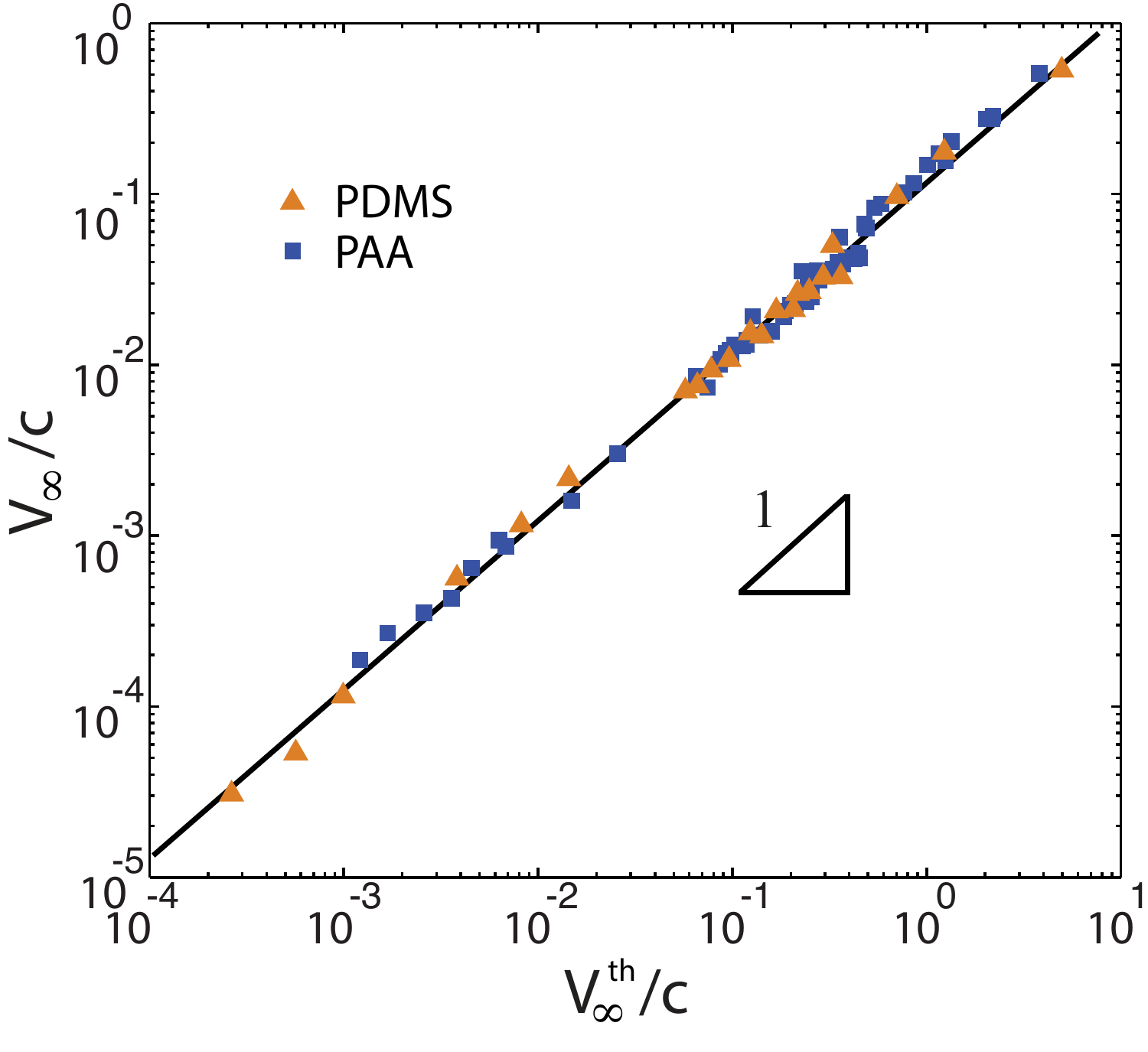}
\end{center}
\caption{\textbf{Master curve for steady sliding speed of an immersed cylinder on a soft substrate.} Time-averaged steady-state sliding speed $V_{\infty}$ (Fig.~\ref{fig2}) as a function of its theoretical prediction $V^{\textrm{th}}_{\infty}$ (Eq.~(\ref{master})). Both axes are scaled by the free fall speed $c=\sqrt{2gR\rho^*/\rho}$. The experiments were performed using either PDMS elastomers (triangles) or PAA hydrogels (squares). Both materials give overlapping results for a range of moduli. The solid line has unit slope.\label{fig3}}

\end{figure}

At steady state, $V=V_{\infty}$, the elastohydrodynamic lift normal to the incline must balance the normal projection of the cylinder weight $\sim \rho^* g R^2 \cos \alpha$, while the tangential gravitational driving power $\sim V_{\infty}\rho^*gR^2 \sin\alpha$ must balance the resisting viscous power $\sim \eta(V_{\infty}/\delta)^2l\delta$. These relations yield the theoretical steady-state sliding speed:
\begin{equation}
\label{master}
V^{\textrm{th}}_{\infty}=A \frac{\rho^* g R^2 \sin \alpha}{\eta} \left(\frac{\rho^* gh \cos \alpha}{G}\right)^{1/5} \left(\frac{\sin \alpha}{\cos\alpha}\right)^{2/5}\ ,
\end{equation}
where $A$ is a dimensionless constant. We note that $V^{\textrm{th}}_\infty$ is the Stokes velocity $\sim \rho^*g R^2\sin \alpha/\eta$ of a particle of size $R$ in an infinite fluid, modified by the effect of elastohydrodynamics; indeed, the steady-state gap reads $\delta_\infty \sim R (\rho^* gh \cos \alpha/G)^{2/5} (\sin \alpha/\cos \alpha)^{4/5}$ and allows us to rewrite the above scaling law in terms of the gap under more general loading conditions. A first check of Eq.~(\ref{master}) can be seen already in Fig.~\ref{fig2} showing evidence of the steady speed for a range of elastic layer thicknesses and moduli; furthermore, the speed is inversely proportional to $G^{1/5}$ in the small deformation limit, i.e. for relatively stiff materials. To test Eq.~(\ref{master}) further, we used cylinders (all of width $12.7$~mm) with radii $R=12.7$, $9.5$, and $6.35$~mm, and densities $\rho=8510$, $2720$, and $2240$~kg/m$^3$ corresponding to brass, aluminium and glass, moving in fluids with a range of viscosities $\eta=0.01$, $0.35$, $1$, $30$, and $100$~Pa.s, along an incline with angle $\alpha \in[10^{\circ}-40^{\circ}]$, that is coated with two different soft materials (PDMS, PAA) of modulus $G \in [8-300]$~kPa (independently measured using a CP$50$ cone-plate geometry in an MCR $501$ Anton Paar rheometer, see SI) and thickness $h \in [100~\mu\textrm{m} - 2~\textrm{mm}]$. In Fig.~\ref{fig3}, we see that the measured sliding speed is in excellent agreement with Eq.~(\ref{master}), over five decades in the scaled speed for both PDMS and PAA materials. Furthermore, the experimental prefactor $A=0.12\pm0.02$ in Eq.~(\ref{master}) is consistent with the theoretical prediction $A\approx0.2$~\cite{Salez2015}.

While Eq.~(\ref{master}) captures the small-deformation regime associated with self-sustained lift and low friction, it cannot explain the decreasing of the sliding speed when the substrate is very soft, as shown in Fig.~\ref{fig2}(b), that is when $\Delta h/h \ge 10\%$ as suggested by laser profilometry. This can be explained by recalling that large deformations would require nonlinear extensions of the present theory~\cite{Skotheim2005}. However, it is worth noting that steady-state sliding is still observed in this regime and the sliding speed remains larger than in the case of a rigid substrate, suggesting that the lift force remains important even in this regime. 

Our study of the motion of a fluid-immersed cylinder on a soft incline shows how self-sustained lift and low friction emerge dynamically due to the asymmetric elastic deformation induced by the lubrication flow in the contact zone, in quantitative agreement with a recent theory~\cite{Salez2015}. The origin of the low friction lies in a self-sustained lift that is due to the viscous-flow-induced fore-aft asymmetry of the deformation profile of the incline, and leads to an elastohydrodynamic analogue of Reynolds slider bearing~\cite{Reynolds1886}. This yields a scaling law for the steady velocity as a function of the fluid viscosity, the elastic modulus and thickness of the substrate, gravity, inclination angle, and the size of the particle, which we corroborate using experiments.  These observations, when combined with substrate poroelasticity~\cite{Skotheim2004}, may partly explain a variety of phenomena that couple flow and deformation at soft interfaces such as long run-out landslides~\cite{Campbell1989} or low friction and wear of synovial joints~\cite{Greene2011}. Our study also points to a simple design principle for the reduction of friction and wear at soft interfaces, by tuning the elastohydrodynamic interaction between soft particles using their shape and {deformation properties}, and might allow us to design fluid suspensions with controllable rheologies. 
\newline

{\bf Acknowledgments}
We thank Shmuel Rubinstein for help with imaging, Jun Chung for discussions, and the MacArthur Foundation (LM) and the Harvard MRSEC DMR-1420570 for partial financial support. We also thank Martin Essink, Anupam Pandey, and Jacco Snoeijer for a careful reanalysis of the data and for interesting discussions.

\pagebreak
\widetext

\renewcommand{\thefigure}{S\arabic{figure}}
\setcounter{figure}{0}
\renewcommand{\theequation}{S\arabic{equation}}
\setcounter{equation}{0}

\section*{Supplementary Information}
\subsection*{Experiments with elliptic cylinders}

\begin{figure}[h!]
\includegraphics[width=0.45\textwidth]{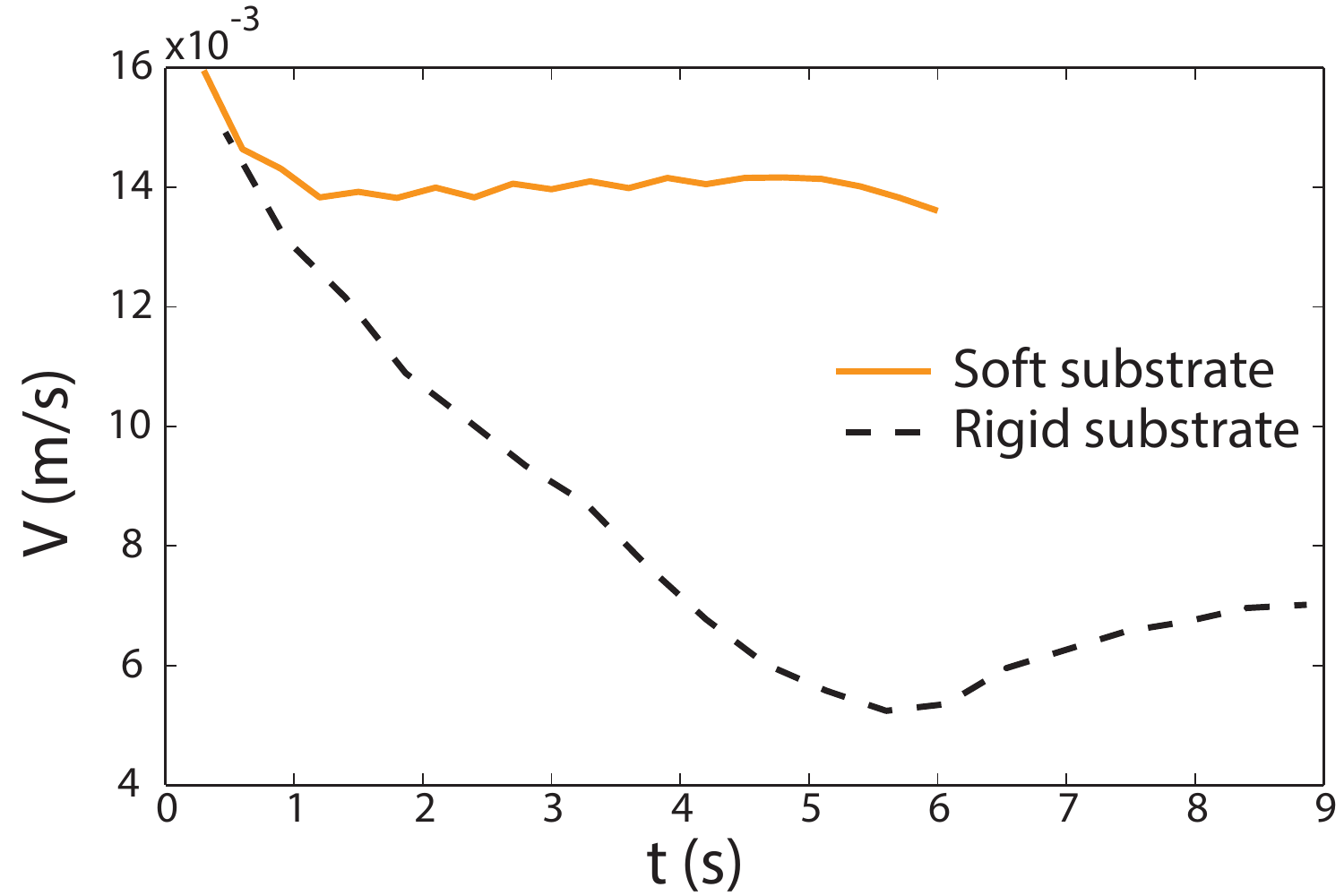}
\includegraphics[width=0.45\textwidth]{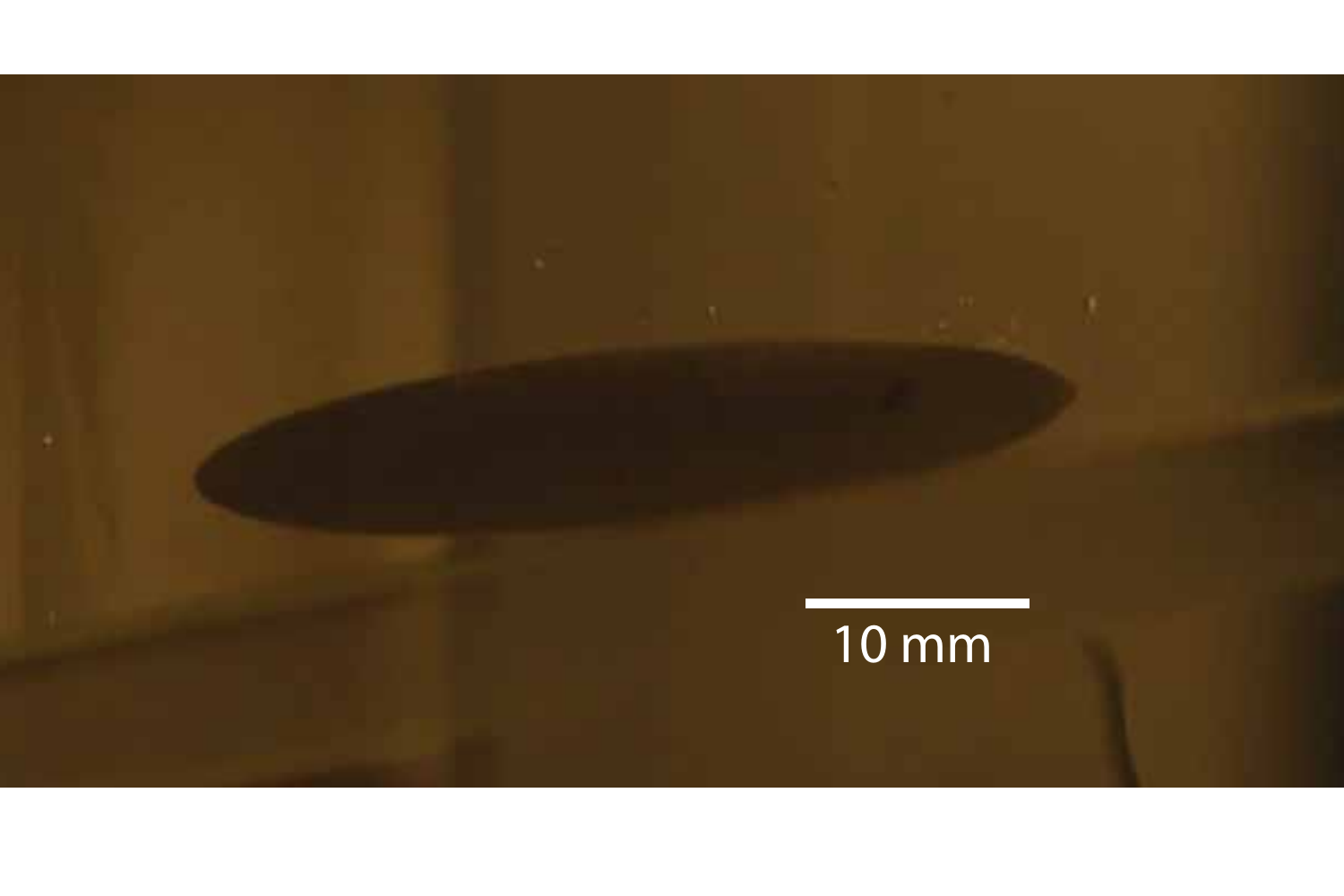}
\caption{\textbf{Experiments with elliptic cylinders.} (left) Sliding speed $V$ as a function of time $t$. (right) Picture of a brass elliptic cylinder during the fall, with mass $24.4$~g, density $\rho=8510$~kg/m$^3$, $H=7.6$~mm, $L=38.1$~mm. The elliptic cylinder falls in a bath of viscosity $\eta=1$~Pa.s near an incline with angle $\alpha=11^{\circ}$. The shear modulus of the soft PDMS coating atop the incline is $G=30$~kPa and its thickness is $h=1.5$~mm. Video M3 is a movie of those experiments.}
\label{ellipse}
\end{figure}

We made experiments with 4 different brass elliptic cylinders of density $\rho=8510$~kg/m$^3$, mass $m=[16.2, 24.24, 27.33, 55.03]$~g, respective major axis $L=[25.4, 38.1, 25.4, 50.8]$~mm and minor axis $H=[7.6, 7.6, 12.7, 12.7]$~mm. The experiments were performed near a substrate of thickness $h=1.5$~mm and shear modulus $G=30$~kPa, tilted at an angle $\alpha=11^{\circ}$, inside an oil bath of viscosity $\eta=1$~Pa.s. They all exhibited the same qualitative behaviour as the one shown in Fig.~\ref{ellipse}: the sliding speed of elliptic cylinders is higher and more stable near a soft substrate than near a rigid glass surface.

\subsection*{Materials}

PDMS and PAA are chosen for the range of moduli they allow to explore with a purely elastic behaviour, and for the simplicity of their fabrication protocol. In order to control the material properties, we performed a rheological study that we present in the following section. All the materials were tested using an Anton Paar (Physica MCR 501) rheometer. The tests were done under an oscillating CP50 cone-plate geometry, with an angular frequency of 10~rad/s and a strain of 0.1\% for PDMS and 0.5\% for PAA. The curing temperature was controlled using an integrated Peltier module. 

\subsubsection*{Polyacrylamide (PAA)}
  
PAA allows to get elastic substrates with shear moduli between $G=0.1$ and $37$~kPa. It is a chemical gel, \textit{i.e.} the monomers are linked by covalent bonds with irreversible curing. The PAA is created by mixing a single solution of monomers and crosslinkers (40\% solution of 37.5:1 acrylamide and bis-acrylamide from Bio-Rad, solution MIX) for the higher moduli, or two separated solutions of acrylamide (4~mol/L, solution A) and bis-acrylamide (0.04 mol/L, solution B) for the lower moduli, with a curing agent in the form of a solution of potassium persulfate (0.2~mol/L, solution C) and a solution  of tetramethylethylenediamin (TMEDA) (0.3~mol/L, solution D) used to create the free radicals.

\begin{figure}[h!]
\includegraphics[width=0.45\textwidth]{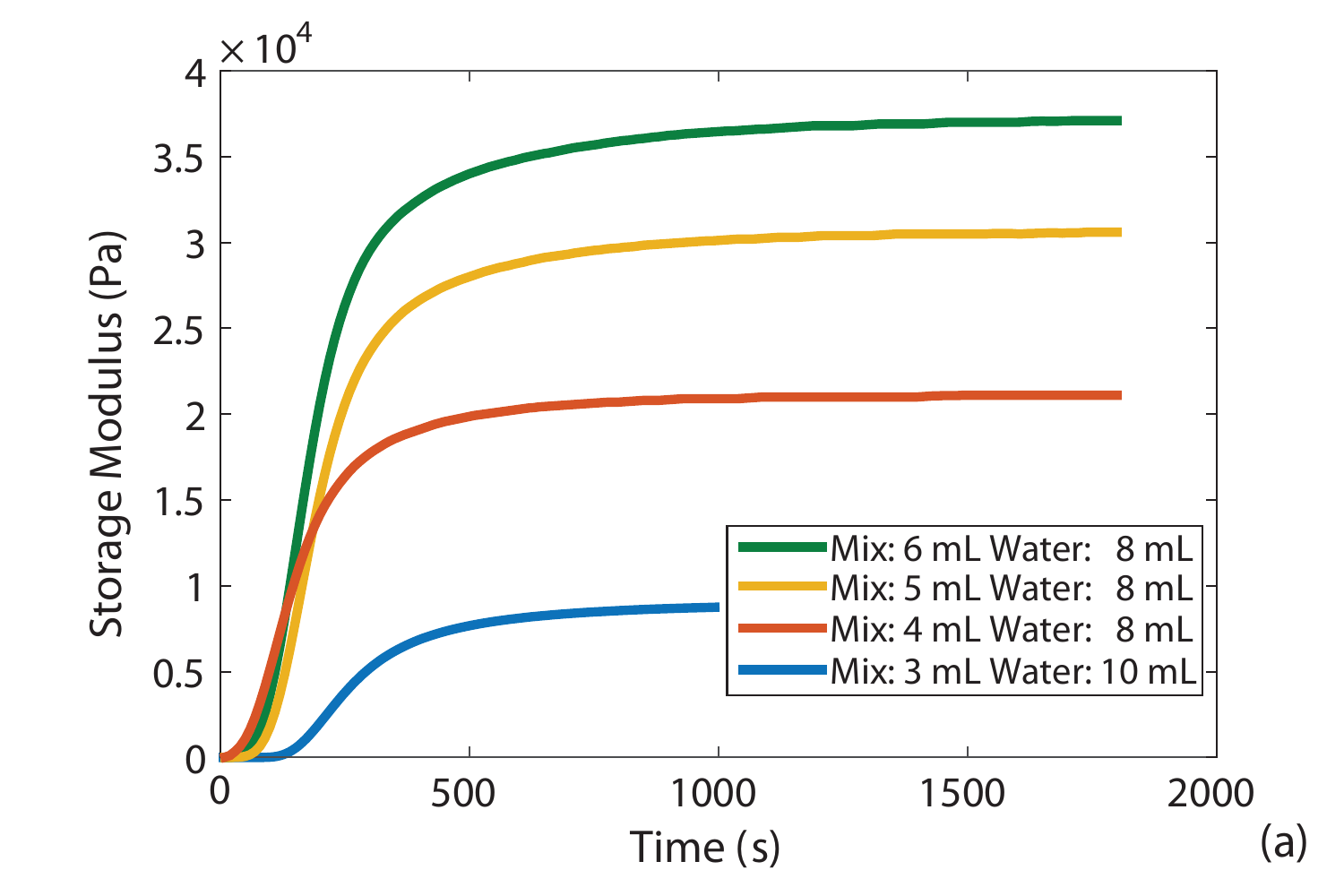}
\includegraphics[width=0.45\textwidth]{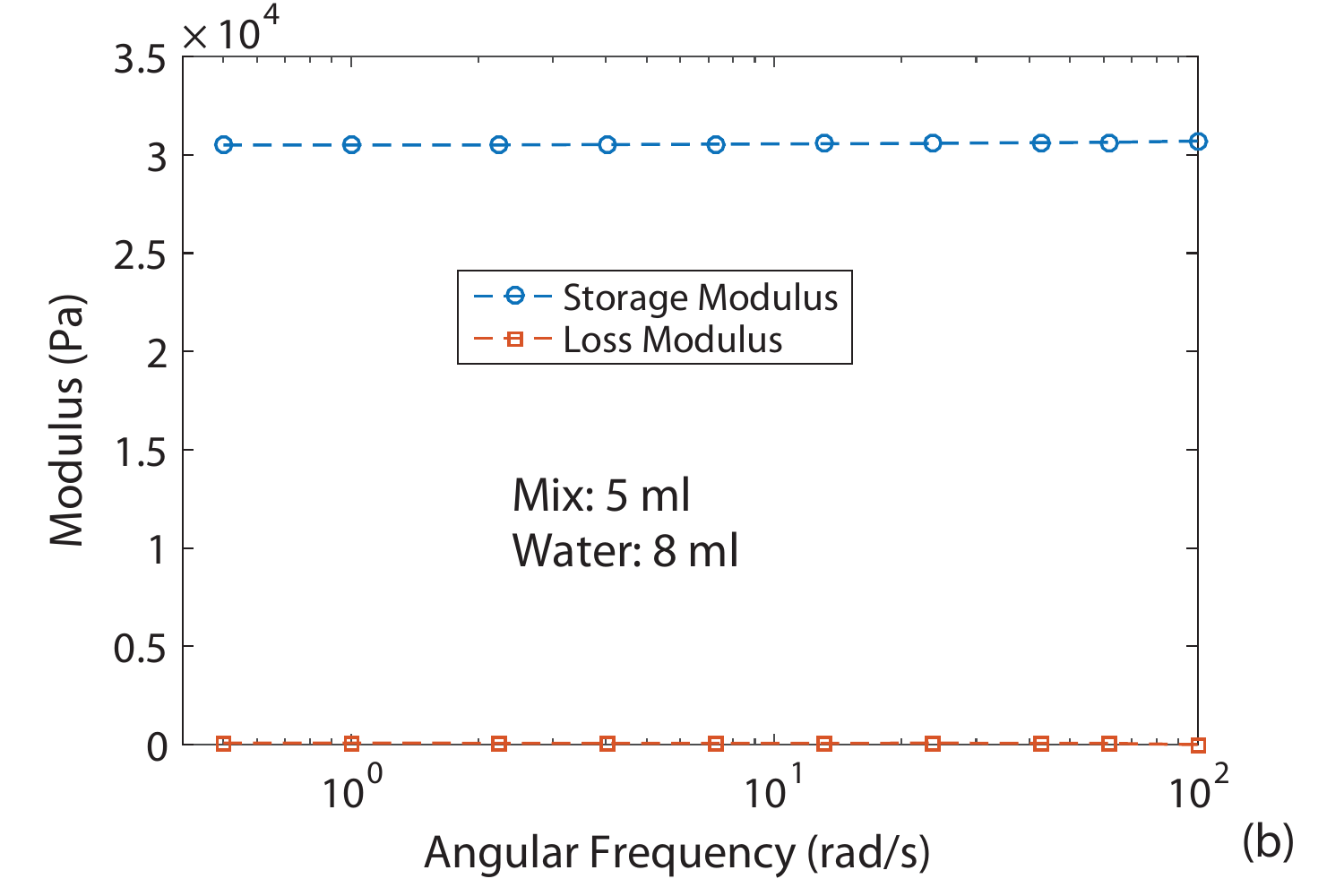}
\caption{\textbf{PAA properties.} (a) Storage modulus as a function of curing time at room temperature, for different volumes of MIX, 0.8~mL of C and 1.2~mL of D. (b) Storage and loss moduli as a function of the angular frequency, after curing. The gel composition is 5~mL of MIX, 8~mL of water, 0.8~mL of C and 1.2~mL of D.}
\label{polyac1}
\end{figure}

To explore a wide range of moduli, we used different concentrations of those solutions and tested the obtained moduli during the curing in the rheometer as previously done in \cite{Saintyves2013}. As shown in Fig.~\ref{polyac1}(a), the storage modulus is stabilising after $\sim20$~min, which provides the time to wait when making the gel layer. For each recipe, at the considered frequencies, the loss modulus is negligible compared to the storage modulus. This storage modulus is also independent from the frequency of the test as shown in Fig.~\ref{polyac1}(b), confirming that the mechanical behaviour is purely elastic.

\subsubsection*{ Polydimethylsiloxane (PDMS)}

PDMS allows to get elastic substrates with shear moduli in the range $G=35$ to $400$~kPa. This material has been studied thoroughly~\cite{0960-1317-24-3-035017}. As for PAA, to develop a set of recipes corresponding to a wide range of moduli, we investigated the mechanical properties of the commercial PDMS Sylgar 184 as a function of the curing protocol and monomer concentration. The Sylgar 184 silicon elastomer is created from two parts: a monomer base (part A) and a curing agent (part B). We control their weight ratio by using a balance. Then, we manually mix the two parts vigorously during $5$~min and put the mix in a vacuum desiccator during $90$~min to remove the trapped bubbles. The degassed PDMS can then be used simultaneously for the rheological test and for the main experiment. When the mixed PDMS is placed in the rheometer cell, the temperature is set to be the same as in the curing oven used for the glass curing cell. 

\begin{figure}[h!]
\includegraphics[width=0.45\textwidth]{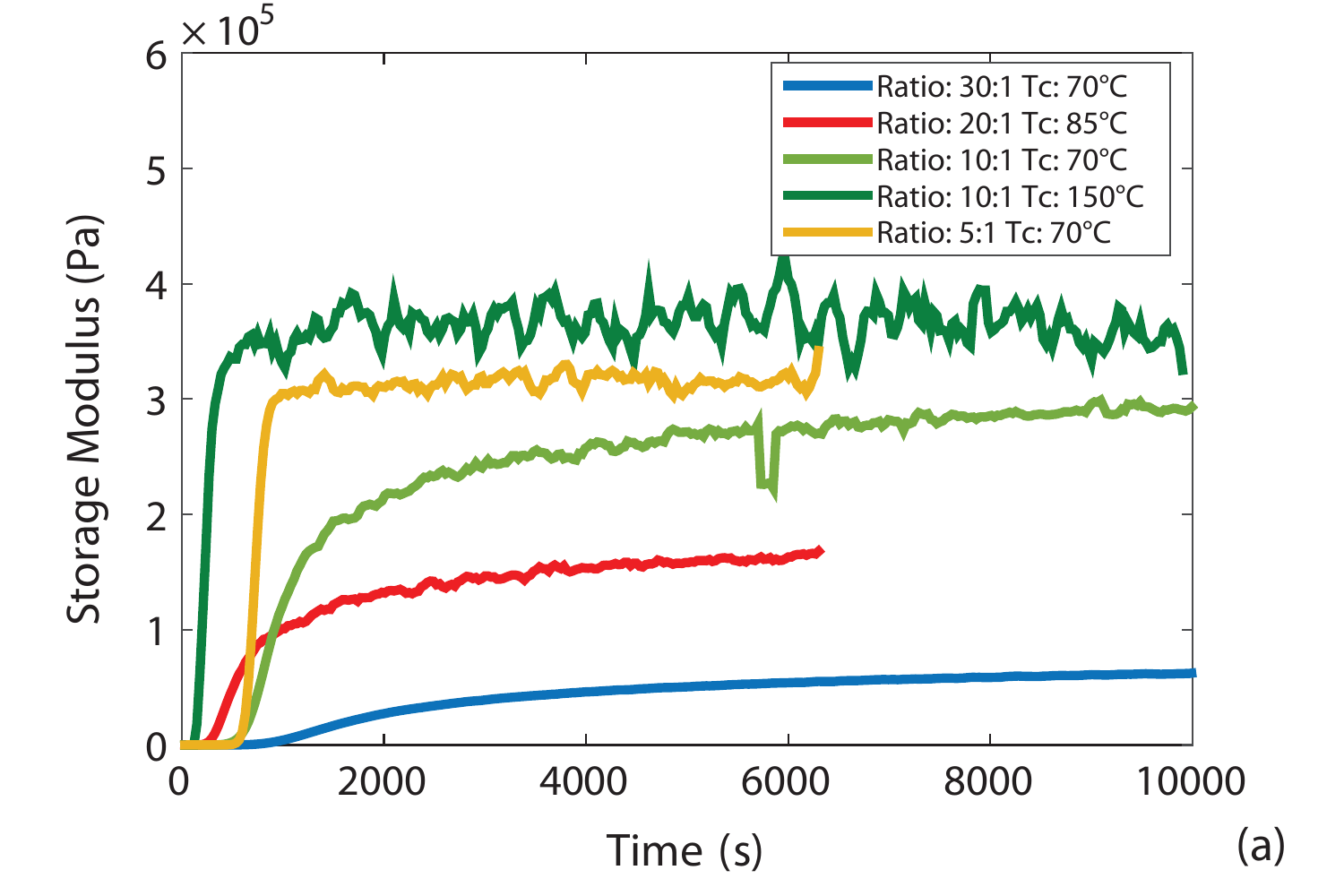}
\includegraphics[width=0.45\textwidth]{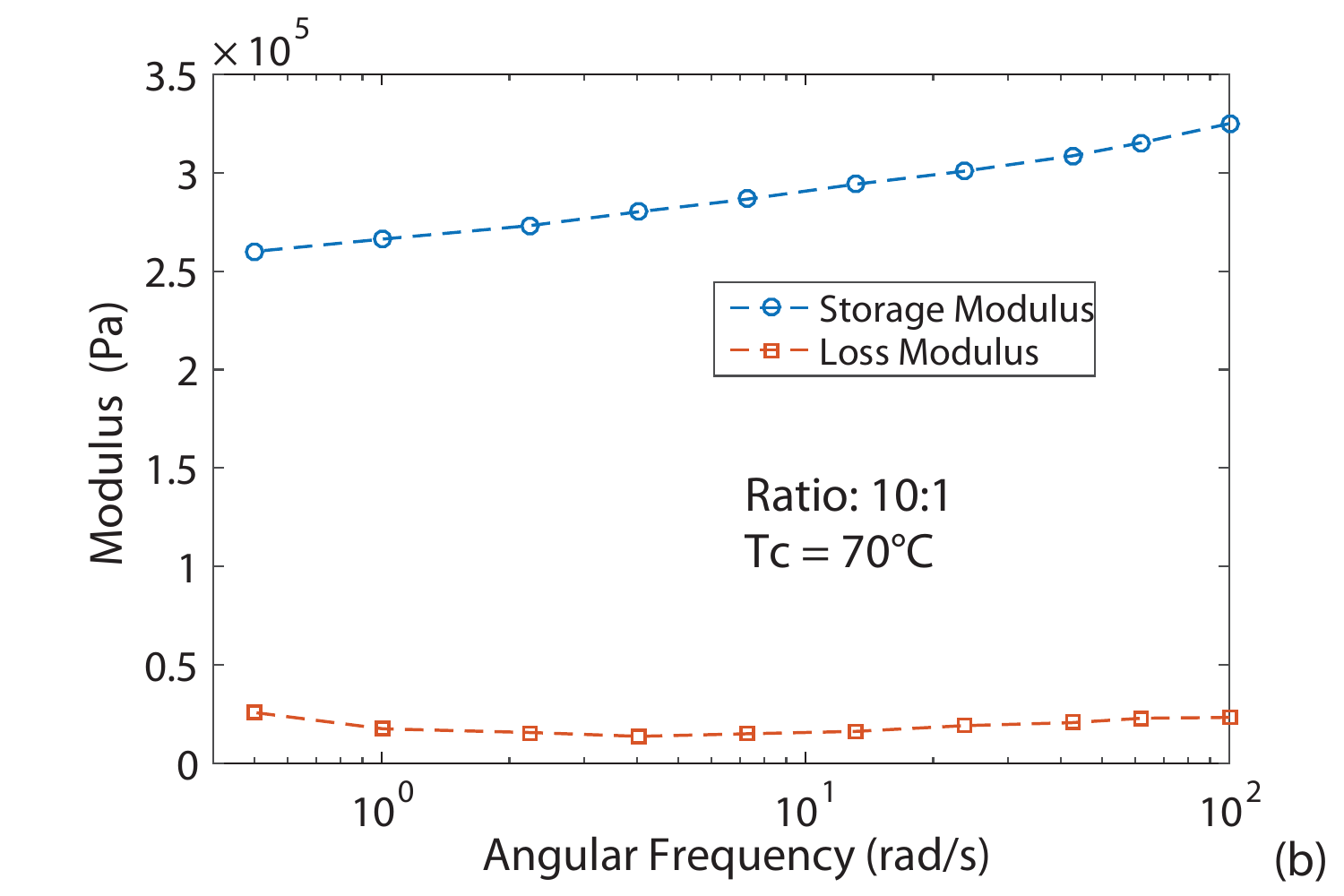}
 \caption{ \textbf{PDMS properties.} (a) Storage modulus as a function of curing time at fixed temperature $T_{\textrm{c}}$, for different ratios of A and B, as indicated. (b) Storage and loss moduli as a function of the angular frequency, after curing. The measurements are done with a 10:1 ratio of A and B and the curing temperature is $T_{\textrm{c}}=70^{\circ}$~C.}
\label{pdms1}
\end{figure}

Figure~\ref{pdms1}(a) shows measurements of the storage modulus over time for different protocols. Characteristic stabilisation times range between $10$ and $180$~min, and provide the times to wait before the curing is complete. Those results are in agreement with observations by ~\cite{0960-1317-24-3-035017}. Figure~\ref{pdms1}(b) shows that the loss modulus is much lower than the storage modulus in the considered range of frequencies. We did not use ratios higher than 30:1, for which the frequency dependency of the modulus might become significant. 

\subsection*{Movie M1}
 \textbf{Deformation profile during the sliding of a cylinder near a soft substrate.} The experimental parameters are: $G=65$~kPa, $h=1.5$~mm, $\eta=1$~Pa.s, $R=12.7$~mm, $\rho=8510$~kg/m$^3$ and $\alpha=11^{\circ}$. 
 
\subsection*{Movie M2}
\textbf{Superposition of cylinders sliding near soft and stiff substrates.} These experiments correspond to a brass cylinder of diameter $12.7$~mm, thickness $12.7$~mm, and density $\rho=8510$~kg/m$^3$. The oil bath viscosity is $\eta=1$~Pa.s and the substrate angle is $\alpha=11^{\circ}$. In the soft case, the substrate thickness is $h=0.6$~mm and the shear modulus is $G=4$~kPa. 
 
\subsection*{Movie M3}
\textbf{Inverse rotation during the sliding along a soft substrate.}  
The falling cylinder is made of brass of density $\rho=8510$~kg/m$^3$ with a diameter of $25.4$~mm and a thickness of $12.7$~mm. The other experimental parameters are $G=16$~kPa, $h=0.2$~mm, $\eta=1$~Pa.s and $\alpha=12.4^{\circ}$.

\subsection*{Movie M4}
\textbf{Oscillations during a fall towards a soft substrate.} The movie speed is slowed down by $10$ times. The falling cylinder is made of brass of density $\rho=8510$~kg/m$^3$ with a diameter of $25.4$~mm and a thickness of $12.7$~mm. The other experimental parameters are $G=31$~kPa, $h=1.5$~mm, $\eta=0.01$~Pa.s and $\alpha=14^{\circ}$.

\subsection*{Movie M5}
\textbf{Superposition of elliptic cylinders sliding near soft and stiff substrates.} Brass elliptic cylinders of density $\rho=8510$~kg/m$^3$, mass $m=24.24$~g, major axis $L=38.1$~mm, and minor axis $H=7.6$~mm. The cylinders fall on an incline at an angle $\alpha=11^{\circ}$ in an oil bath of viscosity $\eta=1$~Pa.s. The faster one corresponds to a fall along a soft substrate of thickness $h=1.5$~mm and shear modulus $G=30$~kPa. The slower one corresponds to a fall along a non-coated rigid glass surface.

\subsection*{Movie M6}
\textbf{Cylinder sliding near a very soft substrate.} The cylinder is made of brass of density $\rho=8510$~kg/m$^3$ with a diameter of $12.7$~mm and a thickness of $12.7$~mm. The other experimental parameters are $G=500$~Pa, $h=0.6$~mm, $\eta=1$~Pa.s and $\alpha=11^{\circ}$.

\end{document}